# Origin of the superconductivity of WTe$_2$ under pressure

Pengchao Lu[1*], Joon-Seok Kim[2*], Jing Yang[5], Hao Gao[1], Juefei Wu[1], Dexi Shao[1], Bin Li[3], Dawei Zhou[1,4], Jian Sun[1]†, Deji Akinwande[2,6], Jung-Fu Lin[5,6,7], Dingyu Xing[1]

[1] National Laboratory of Solid State Microstructures, School of Physics and Collaborative Innovation Center of Advanced Microstructures, Nanjing University, Nanjing 210093, China.

[2] Microelectronics Research Center, Department of Electrical and Computer Engineering, The University of Texas at Austin, Austin, TX 78758, USA

[3] College of Science, Nanjing University of Posts and Telecommunications, Nanjing 210023, China

[4] College of Physics and Electronic Engineering, Nanyang Normal University, Nanyang 473061, China

[5] Department of Geological Sciences, Jackson School of Geosciences, The University of Texas at Austin, Austin, TX 78712, USA

[6] Texas Materials Institute, The University of Texas at Austin, Austin, TX 78712, USA

[7] Center for High Pressure Science & Technology Advanced Research (HPSTAR), Shanghai 201203, People's Republic of China

*These authors contributed equally to this work. †Correspondence and requests for materials should be addressed to J.S. (e-mail: jiansun@nju.edu.cn)

# Abstract

Tungsten ditelluride (WTe$_2$) has attracted significant attention due to its interesting electronic properties, such as the unsaturated magnetoresistance and superconductivity. Recently, it has been proposed to be a new type of Weyl semimetal, which is distinguished from other transition metal dichalcogenides (TMDs) from a topological prospective. Here, we study the structure of WTe$_2$ under pressure with a crystal structure prediction and ab initio calculations combined with high pressure synchrotron X-ray diffraction and Raman spectroscopy measurements. We find that the ambient orthorhombic structure (Td) transforms into a monoclinic structure (1T') at around 4-5 GPa. As the transition pressure is very close to the critical point in recent high-pressure electrical transport measurements, the emergence of superconductivity in WTe$_2$ under pressure is attributed to the Td-1T' structure phase transition, which associates with a sliding mechanism of the TMD layers and results in a shorter Te-Te interlayer distance compared to the intralayer ones. These results highlight the critical role of the interlayer stacking and chalcogen interactions on the electronic and superconducting properties of multilayered TMDs under hydrostatic strain environments.

2D transition-metal dichalcogenides (TMDs) have recently become one of the most extensively investigated system due to their unique properties including tunable electronic and transport properties, high mechanical flexibility, as well as the ability to be assembled into layered heterostructures using vertical fabrications technique [1]. These TMDs materials with a general chemical formula $MX_2$ consists of a layer of transition metal atoms (M=Mo, W, etc.) in the middle that is covalently sandwiched by two layers of chalcogenide atoms (X=S, Se, or Te). The X-M-X layers bond with other layers via weak van der Waals forces. The number of X-M-X layers and the arrangements of atoms within the layers of TMDS can significantly affect their electronic and transport properties including their bandgap structures. For instance, $MoS_2$ has a sizable band gap of 1.29eV to 1.90eV, changing from an indirect band gap in bulk to a direct one in single layers [2,3,4]. More importantly, thin-film $MoS_2$ transistors exhibit remarkably high mechanical flexibility [5,6], high ON/OFF ratio, carrier mobility, and low operating voltage, promising to become high-performance logic and photo detector devices.

Among the Group VIb element TMDs, tungsten ditelluride ($WTe_2$) has been recently reported to possess a number of extraordinary physical properties, which makes it a very interesting system for research. For instance, a large unsaturated magnetoresistance effect has been observed under a magnetic field of up to 60 Tesla at cryogenic temperatures below 4.5K [7]. The electrical resistance along *a* axis of $WTe_2$ lattice increases significantly when a magnetic field is applied perpendicularly to Te layers (along *c* axis). As demonstrated in the framework of semi-classical two-band model theory combined with first-principles calculations [8] and angle-resolved quantum oscillations measurements [9], the enhanced resistance is attributed to the symmetric balance between electron and hole pockets on the Fermi surface. Pressure-induced superconductivity in WTe$_2$ was observed by two different groups [10,11] very recently, the critical pressure for the emergence of the superconductivity varies from 2.5 GPa to 10.5 GPa. The critical temperature (Tc) reaches a maximum of around 6-7 K at certain pressure and decreases with compression to form a dome-like diagram. [10, 11] Furthermore, a quantum phase transition is proposed from high-pressure electrical resistance measurements and the pressure-dependent Hall coefficient with a sign changing from the positive to the negative at around 10.5 GPa [11]. The soft modes in the calculated phonon spectra of Td under high pressure indicate a possible structural phase transition. [10] However, clear structural phase transition has not been thoroughly explored along with the emergence of superconductivity [10, 11], the structure of the superconducting phase and the origin of the superconductivity in WTe$_2$ under pressure still remain to be elucidated.

A new type of Weyl semimetal (WSM) (Type II) in WTe$_2$ is also recently proposed by Soluyanov *et al.* [12] in which the Weyl points locating at 0.052 eV and 0.058 eV above the Fermi level exist at the boundary of electron and hole pockets, rather than at the point-like Fermi surface in traditional WSM (Type I) systems. Motivated by this prediction, Td-MoTe$_2$ is also predicted to be a type II WSM, but the Weyl points are more close to Fermi level (about 6 meV) and the spacing between each pair of the Weyl points is six times larger than that of WTe$_2$ [13]. As a clean way to adjust the lattice constants and so to the interactions, applied pressure has been used to produce new topological phases of materials [14]. Therefore, whether pressure could change the topological properties of WTe$_2$ is also becoming an interesting question.

Tungsten ditelluride (WTe$_2$) crystallizes in the orthorhombic structure with a distorted octahedral coordination (Td) (space group Pmn21, No.31) under ambient conditions [15], which is uniquely different from the typical trigonal prismatic structure in other TMDs, such as the 2H-MoS$_2$, 2H-WSe$_2$ (space group P63/mmc, No. 194) [16] or the monoclinic structure 1T'-MoTe$_2$ (space group P21/m, No. 11) [15]. Since the crystal structure of WTe$_2$ is distinguishable from other TMDs at ambient conditions, it is reasonable to assume that the crystal structure of WTe2 at high pressures can play a significant role in the origin of these aforementioned unique physical properties. However, despite its significance, high-pressure phase stability of WTe$_2$ remains largely unknown and has hindered our understanding of the fundamental physics of the TMDs system at extreme environments.

To give new insights on the aforementioned scientific issues, we explore the structures of WTe$_2$ under pressure up to 30 GPa, with a combination of ab initio calculations and experiments including high pressure synchrotron powder X-ray diffraction and Raman measurements. Ab initio random structure searching techniques (AIRSS) [17, 18] is used for crystal structure predictions, ab initio calculations are performed with VASP code [19]. The van der Waals (vdW) interaction is carefully taken into account using vdW-DF corrections of Langreth and Lundqvist *et al*. [20] together with optB88 functionals [21, 22, 23]. The electron-phonon coupling (EPC) calculation implemented in Quantum-ESPRESSO program [24] is used to estimate the critical temperature. Details of the methods can be found in Supplemental Material. We have predicted two high-pressure phases in WTe$_2$ with P2$_1$/m and P6$_3$/mmc symmetry, which resemble the most common 1T' and 2H structure in TMDs, respectively. Moreover, we synthesized the 1T' phase in the high pressure XRD and Raman experiments, and resolved its structure. The Td to 1T' transition pressure is close to the pressure where the superconductivity occurs [10, 11], making it likely that the superconducting WTe$_2$ is actually in the 1T' phase.

The crystallographic structures of Td, 1T' and 2H phases are shown in Fig. 1 (a), (b). To compare 2H with Td and 1T', the hexagonal 2H lattice is converted to orthorhombic lattice through building a $1 \times \sqrt{3}$ supercell in xy plane. In the Td phase, the tungsten atoms are sandwiched between two layers of tellurium atoms in which one layer is rotated 180 degrees with respect to the other, forming the W-Te$_6$ octahedral coordination. These layers in the Td phase exhibit zigzag-like structure, whereas the high-pressure 2H structure only contains flat layers. 1T' structure is very similar to Td such that it can be constructed from Td phase with shear strain, which introduces an inversion center. Viewing the structure from the out-of-plane direction, tungsten and tellurium atoms are packed in a honeycomb crystal lattice for the 2H structure, similar to graphene or h-BN, but the Td structure displays a rather complicated atomic arrangement as well as 1T' phase. The calculated enthalpy-pressure (ΔH-P) and volume-pressure (V-P) curves of WTe$_2$ are shown in Fig 1. (c), which clearly show two structural phase transitions, from Td to 1T' at around 5 GPa and from 1T' to 2H phase at about 10 GPa. Td structure is found to be thermodynamically stable at ambient conditions, in agreement with the experimental observation [15]. However, the relative enthalpy difference between Td and 1T' phase is very small, less than 0.5meV per formula unit, at ambient pressure. Moreover, calculated lattice parameters of Td and 1T' phase are very close, as well as their volumes. The 2H phase is much more thermodynamically stable under high pressure, with lower enthalpy than that of 1T' phase by 150 meV/f.u. at 25GPa. As 2H structure with flat layers is more compact than the zigzag-like 1T' structure, there is a sudden drop in volume for 1T'-2H transition. Compared with some other typical TMDs with flat layers, WTe$_2$ has un-flat layers, in which pressure acts as some kind of iron to smooth the layers and provide a "compressive strain" to result in higher symmetry and tune the structure of TMDs as well as its electronic properties. This can be further justified by comparing with $WS_2$, which has stable 2H phase in ambient condition. $WS_2$ does not undergo any structural phase transition up to 60 GPa [25] nor abrupt metallization.

These octahedrally-coordinated W-Te layers are held together by van der Waals interactions such that the dispersion correction is essential for obtaining proper interlayer spacing by calculations. As the lattice parameters and volumes at 0GPa listed in Supplemental Material Table S1, normal GGA-PBE functional gives quite large error as compared to the experimental values, while the calculations based on optB88 functional together with vdW-DF correction, as we used in this work, gives error less than 1.0%.

Dynamical stability of the 1T'- and 2H- $WTe_2$ structure is investigated by calculating their phonon dispersion relations, shown in Supplemental Material Fig. S3. It is confirmed that they are dynamically stable at 0 and 15 GPa. Moreover, the phonon spectra at 0 GPa do not exhibit modes with negative frequencies, indicating that the they should be recoverable in decompression to ambient pressure if they could be synthesized experimentally. It seems that two phonon modes near the A, C and E point goes soft in 1T'-$WTe_2$ at 5GPa.

High pressure X-ray powder diffraction experiment has confirmed the Td-1T' transition. Experimental XRD spectra and simulated XRD patterns of these structures are shown in Fig. 2 (a), (b). The (011) and (113) peak splitting in 1T' phase are observed, which can clearly distinguish 1T' phase from Td phase. The signal of 1T' phase first appears at around 4 GPa, which agrees with the theoretical prediction very well. As increasing pressure with help of external heating to ~350K, the 1T' signal enhances, and finally dominates at higher pressure/temperature (see Fig. 2(a)). A representative experimental XRD spectrum of $WTe_2$ is compared with theoretical prediction of 1T' and Td phase, which resembles 1T' phase in overall range of the spectra (Fig. 2b). More details can be found in Supplemental Material.

There are thirty-three irreducible representations of the optical phonons in Td-WTe$_2$ at Gamma point for the C$_{2v}$ (mm2) group: $\Gamma_{optic}$ = 11A$_1$ + 6A$_2$ + 5B$_1$ + 11B$_2$, which are all Raman active. Seven of them are observed in laser spectroscopic experiments for the Td structure under ambient pressure [26,27] including A$_1$ (78.9 cm$^{-1}$), A$_2$ (88.4 cm$^{-1}$), A$_2$ (109.9 cm$^{-1}$), A$_1$ (114.6 cm$^{-1}$), A$_1$ (129.9 cm$^{-1}$), A$_1$ (160.6 cm$^{-1}$), and A$_1$ (207.7 cm$^{-1}$) [26]. For the 1T' structure with the point group C$_{2h}$ (2/m), there are thirty-three optic modes at Gamma point as well: $\Gamma_{optic}$ = 12A$_g$ + 5A$_u$ + 6B$_g$ + 10B$_u$. Eighteen of them, 12A$_g$ + 6B$_g$, are Raman active modes. The evolution of these modes under high pressure is shown in Fig.2 (c), together with experimental Raman spectra from Ref. [26]. Six Raman peaks measured in ambient pressure, denoted as P1, P2, P3, P5, P6, and P7, undergoes blueshift as pressure increases, where the overall trend shows good match with calculated results up to 30 GPa. The vanishing of P3 and occurrence of P4 with higher frequency above 15 GPa indicate that Td-1T' transition has accomplished (see Fig. 2(c) and Fig. 2(d)). As shown in Fig. 2(e), FWHM of P6 increases dramatically after ~15GPa, meanwhile P7 also experiences rapid broadening after 15GPa. Combined with theoretical prediction, we attribute the widening of P6 and P7 to the appearance of Raman peaks from 1T' phase. The absence of one Raman active mode of 1T' phase (between P5 and P6) may be attributed to its weak intensity.

Analysis of the calculated band structures and Fermi surfaces suggests 1T'-WTe$_2$ to be a semimetal, which is resulted from electron and hole pockets along the Γ-Y direction, shown in Fig. 3. The density of states near Fermi energy is mainly from W 5d and Te 5p orbitals. There are three bands near Fermi level, two electron-like bands and one hole-like band, composing the Fermi surface. The shape of Fermi surface is very sensitive to the pressure, and size of electron and hole pockets increases consequently with pressure increasing, indicating that 1T'-$WTe_2$ undergoes a Lifshitz-like, or so-called electronic topological phase transition under pressure. In fact, band structures and Fermi surface of 1T' phase are very similar to those in Td phase [7], including perfect balance between electron and hole pockets, which may bring large unsaturated magnetoresistance effect in 1T'-$WTe_2$ system as well. In Td phase, spin degeneracy will be broken by the spin-orbital coupling due to absence of inversion symmetry, while in the 1T' phase containing an inversion center, spin degeneracy will remain. Thus the Weyl fermions in the ambient Td phase should disappear after the structure phase transition. In previous work [10,11], it was proposed that superconductivity found in $WTe_2$ emerges from a suppressed large magnetoresistance (LMR) state under high pressure. The critical pressure of superconductivity around 2.5GPa is close to the Td-1T' transition pressure from this work, together with the fact that the Td structure has negative frequencies under high pressure, we believe superconductivity in $WTe_2$ emerges from the 1T' phase rather than the Td phase. Lifshitz transition, which can be seen in Fig. 3 where the topology of Fermi surface develops electron and hole pockets, is reported to coincide with appearance of superconductivity in iron/transition-metal based superconductor [28].

In order to get further understanding of the superconductivity mechanism, we performed ab-initio calculations for the electron-phonon coupling (EPC) of 1T'-$WTe_2$, in the frame work of density functional perturbation theory [29]. The superconducting transition temperature can be estimated by Allen and Dynes modified McMillan formula [30]: $T_c = \frac{\langle \omega \rangle}{1.20} \exp(-\frac{1.04(1+\lambda)}{\lambda - \mu^*(1+0.62\lambda)})$, where the parameter λ is a dimensionless measure of the Eliashberg electron-phonon spectral function $\alpha^2 F(\omega)$, expressed as $\lambda = 2 \int_0^\infty d\omega\, \alpha^2 F(\omega)/\omega$. Using a common value of $\mu^* = 0.1$,

The $T_c$ in 1T'-WTe$_2$ at 10 GPa is estimated to be 2~4 K with different broadening, which agrees well with the measurements. [10,11] Theoretical phonon dispersions of 1T'-WTe$_2$ at 10 GPa are shown in Fig. 4(a). It seems that the low frequency phonon modes indeed have good contribution to the EPC coupling. One TA vibrational mode (B$_u$ mode) near A point is shown in Fig. 4(b), which is mainly contributed by Te-Te interlayer vibrations. Fig. 4(c) shows the calculated interlayer Te-Te distance and W-Te bond length within the layer of 1T'-WTe$_2$ as a function of pressure up to 10 GPa. The Te-Te distance exhibits an obvious variation under pressure while the intralayer W-Te bond length is almost not changed. This implies that the effect of compression is mostly interlayered, thanks to the unique anisotropy of van der Waals structure. The slope of the variation of the Te-Te distance exhibits two linear regions: a rapid decrease with increasing pressure until around 4 GPa and a slow decrease with further increase in pressure, which suggests an irregular point with compression. Add on the facts of the pressure-induced superconductivity in ZrTe$_5$ [31] and MoTe$_2$ [32], the softening of the interlayer Te-Te vibration modes, resulting from compression of zig-zag TMD layers, has reasonable correlations with the emergence of superconductivity in TMDs.

To answer the question of that why the 2H phase has not yet been observed in previous and this high-pressure experiments, we calculated the transition barriers for Td-1T' and 1T'-2H transitions using VC-NEB method [33]. We suppose that there exists a high transition barrier between 1T' (or Td) and 2H phase. As shown in Supplemental Material Fig. S4, due to the similarity between Td and 1T' structure, the transition path can be simply considered as shear strain along b axis. As shear strain does not break any W-Te bond, the barrier between Td and 1T' phase at 5 GPa is quite small, less than 1 meV/atom. For the 1T'-2H transition, as 1T' and 2H structures belong to different stackings, a smooth transition path with no bond broken does not exist. To minimize the number of breaking bonds, we fabricate a simple transition path based on in-plane glide. There are four Te atomic layers in 1T' primitive cell, two next-nearest ones of them are made glide along b axis in opposite directions, these two glides are carry out in different stages to decrease the energy barrier, which results in two barriers and one intermediate low-energy state. The transition barrier between 1T' and 2H phases is estimated to be around 160 meV/atom at pressure of 10 GPa, such high barrier is very difficult to overcome and may be responsible for the missing of 2H-WTe$_2$ phase in the experiments. As forming alloyed MoTe$_2$ −WTe$_2$ monolayers can decrease critical temperatures of T'-H transition efficiently [34], introducing Mo-doping may scale down the transition pressure of 1T'-2H transition in bulk WTe$_2$.

In summary, using crystal structural prediction technique and first-principles calculations, we predict that WTe$_2$ undergoes two structural phase transition, from Td (Pmn2$_1$) to 1T' (P2$_1$/m) at around 4-5 GPa, and then onto 2H (P6$_3$/mmc) phase at around 10 GPa. High-pressure XRD and Raman measurements confirm the Td-1T' transition and give consistent critical pressure of around 4-5 GPa. Since the transition pressure is very close to the emergence pressure of superconductivity and that the irregular points in the Te-Te distance variations and modes softening in the phonon spectra occur across the transition, we thus attribute the pressure-induced superconductivity in WTe$_2$ to Td-1T' structure transition. Based on transition path simulations and energy barrier estimates, we find that the barrier for the Td-1T' transition is very low and the one for the 1T'-2H transition is very high. The latter is likely the reason why the 2H phase has so far not been observed experimentally. These results show pressure not only influences the stacking but also the building blocks of TMDs, which determines the electronic structure and properties of layered materials. This study helps to facilitate a comprehensive understanding of the rich electronic features of $WTe_2$ amongst TMDs materials.

**Acknowledgements**

We acknowledge the financial support from the National Key Projects for Basic Research of China (Grant Nos: 2015CB921202), the National Natural Science Foundation of China (Grant Nos: 51372112 and 11574133), NSF Jiangsu province (No. BK20150012), the Fundamental Research Funds for the Central Universities and Special Program for Applied Research on Super Computation of the NSFC-Guangdong Joint Fund (the second phase). D.A acknowledges support from DTRA and ARO. JFL acknowledges support from HPSTAR Program. High-pressure XRD experiments were conducted at GSECARS of the Advanced Photon Source. The authors thank J. Yang and V. Prakapenka for their assist in the experiments. Part of the calculations were performed on the supercomputer in the High Performance Computing Center of Nanjing University.

# Figures

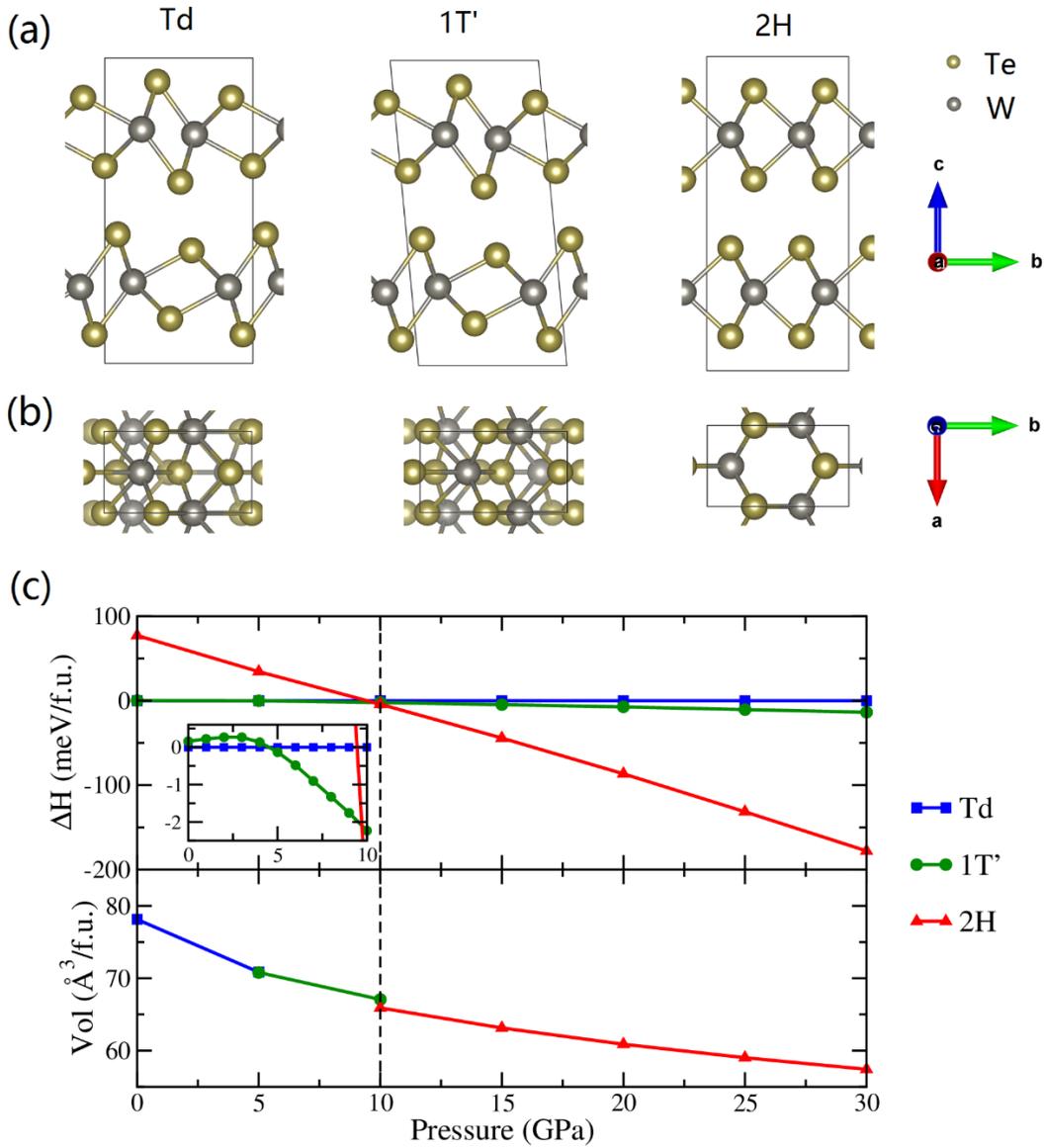

Figure 1. (a) and (b) Schematic representations of Td, 1T' and 2H crystal structures of WTe$_2$ along *a* axis and *c* axis, respectively. To compare 2H with Td and 1T', the hexagonal 2H lattice is converted to orthorhombic lattice by building a $1 \times \sqrt{3}$ supercell in the xy plane. (c) Relative enthalpy difference between 2H and Td phases as a function of pressure (top panel) and their pressure-volume curves (bottom panel). Theoretical calculations indicate two structural phase transitions, from Td to 1T' at around 5 GPa and from 1T' to 2H phase at about 10 GPa, respectively.

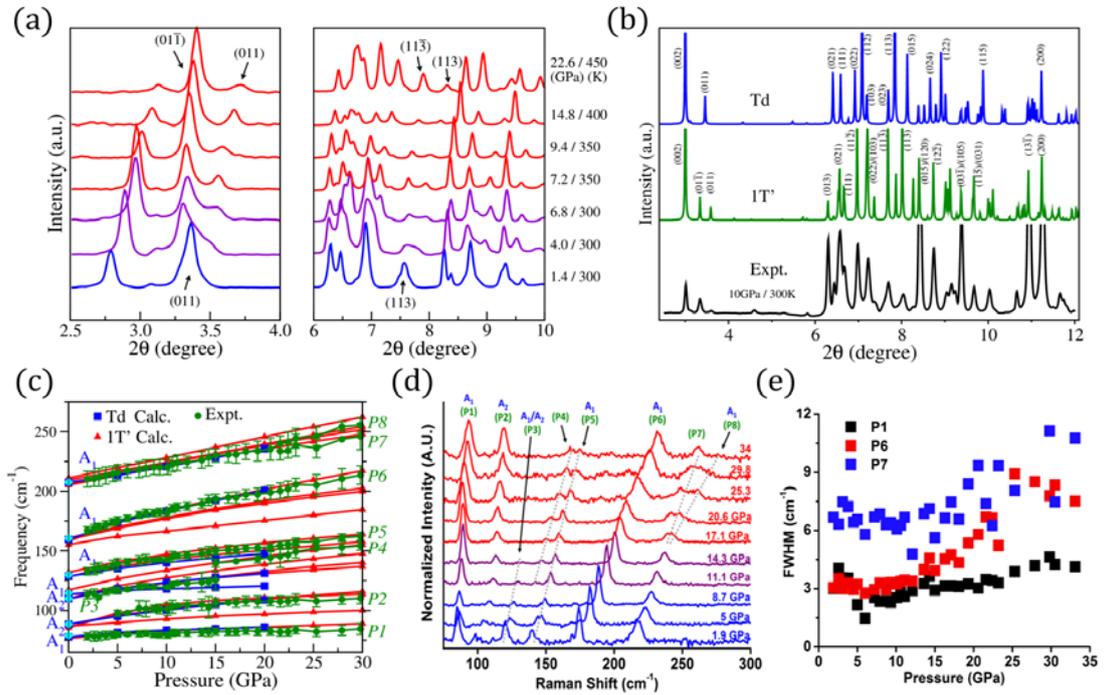

Figure 2. (a) Experimental XRD spectra under high pressure. (011) and (113) peaks clearly shows splitting, which is an evidence of phase transition from Td to 1T' as pressure/temperature increases. (b) Simulated XRD patterns of the optimized structures at 10 GPa, compared with experimental values at 10GPa / 300 K. The wavelength is 0.3344Å. The (011) and (113) peak splitting in 1T' phase are observed, which can distinguish 1T' phase from Td phase. (c) Raman modes of Td- and 1T'- $WTe_2$ under high pressure, compared with experimental values. Calculated results are marked as blue squares for Td phase and red triangles for 1T' phase, respectively. Experimental values under ambient pressure taken from Ref. [26] are marked as cyan diamonds, which agrees well with our calculated results. Overall trend of Raman modes matches well with the experimental data up to 30 GPa, including discontinuity of Td phase mode (P3). (d) Experimental Raman spectra of $WTe_2$ under pressure. Blue region denotes Td-$WTe_2$ phase, purple region denotes the occurrence of Td-1T' transition, red region denotes the 1T' phase. Dashed lines are guided line of the peaks, showing discontinuity of P3 and P4, and phonon splitting of P8 from P7. (e) FWHM of P1, P6, and P7, indicating dramatic broadening of P6 and P7 at pressure >15 GPa. This implies possible rising of Raman modes in 1T' phase, as predicted from theoretical calculation.

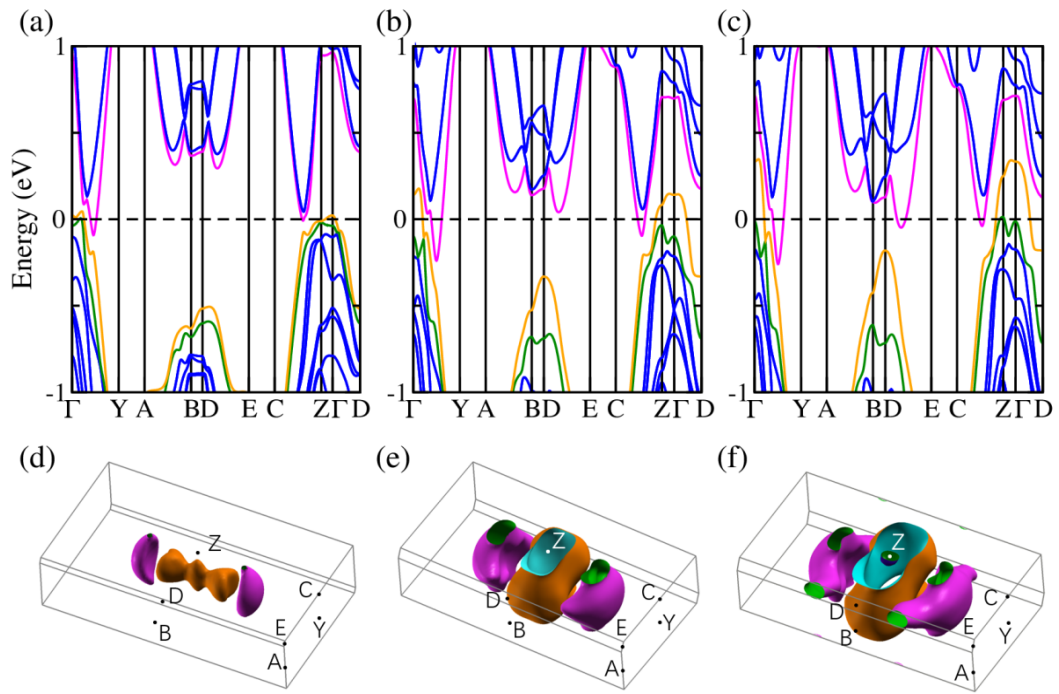

Figure 3. Calculated band structures and Fermi Surfaces of 1T'-WTe$_2$ with spin-orbital coupling (SOC) at 0GPa (a),(d), 5GPa (b),(e), and 10GPa (c),(f). Three bands near Fermi level, two electron-like bands and one hole-like band, are shown in green, orange and magenta respectively, while other bands are in blue. Due to the coexistence of time-reversal symmetry (TRS) and inversion symmetry, the SOC will not break spin degeneracy in 1T'-WTe$_2$.

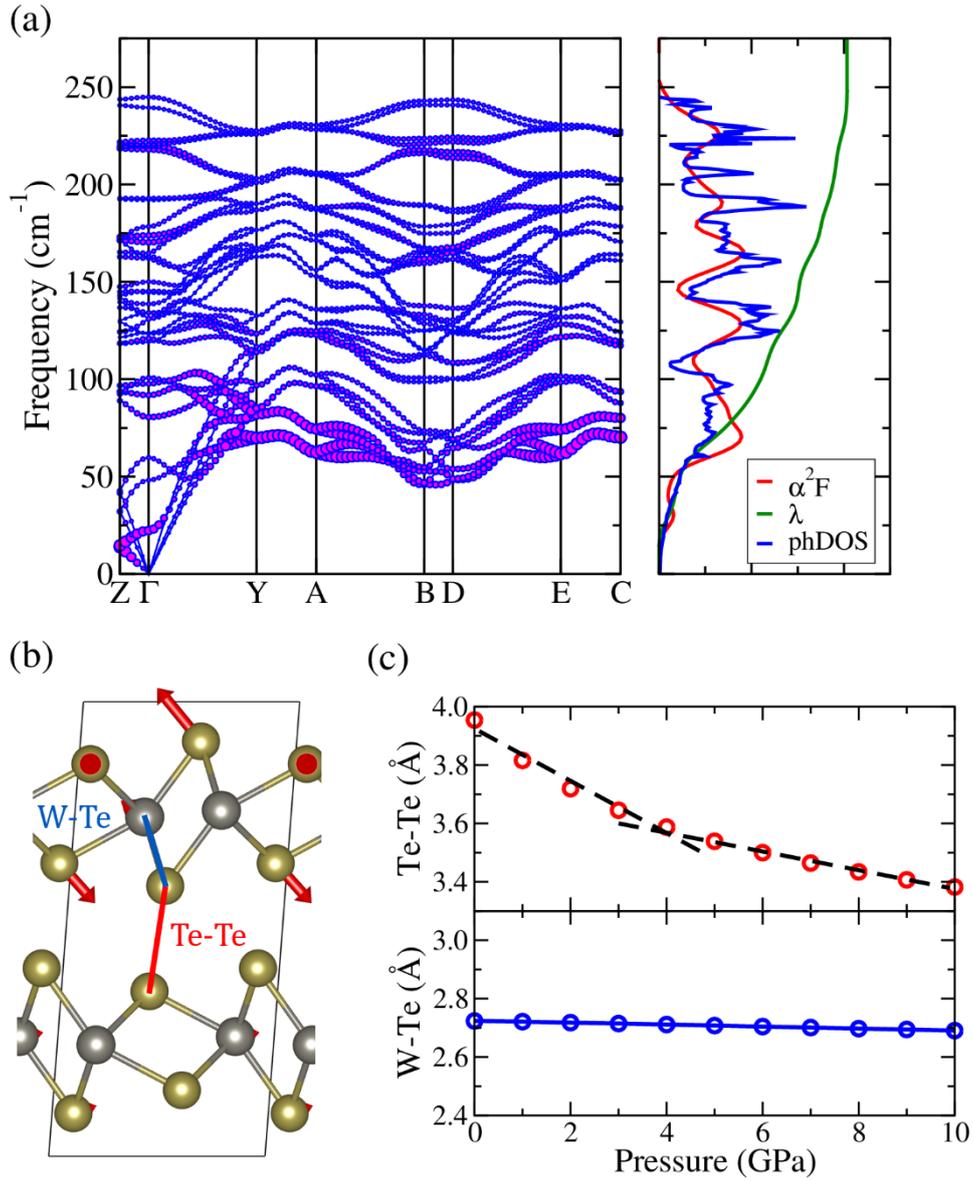

Figure 4. (a) Theoretical phonon dispersions of 1T'-WTe$_2$ at 10 GPa. The size of the dots represents density of electron-phonon spectral function $\alpha^2 F(\omega)$. Phonon DOS and the electron-phonon integral λ(ω) are also shown on the right panel. (b) Schematic representations of 1T'-WTe$_2$ crystal structures along a axis. Red arrows represent one TA vibrational mode (B$_u$ symmetry) near A point. (c) Theoretical Te-Te bond length (top panel) and W-Te bond length (bottom panel) of 1T'-WTe$_2$ as a function of pressure. The Te-Te bond length shows a rapid decrease with increasing pressure until around 4 GPa and a slow decrease with further increase in pressure. The dashed lines fit for two phases, i.e., the normal 1T' phase and the collapsed 1T', in two pressure region.

# Supplemental Material

**Methods**

**Ab initio calculations.** We use ab initio random structure searching (AIRSS) [1,2] method in the frame of density-functional theory (DFT) to find low-enthalpy structures of WTe$_2$, with system sizes up to 8 formula units per simulation cell during the search. Crystal structure optimization calculations are performed by The Vienna Ab initio simulation package (VASP) [3], with Perdew-Burke-Ernzerhof (PBE) [4] generalized gradient approximation (GGA) exchange-correlation density functional and optB88 functional [5,6,7] together with vdW-DF corrections of Langreth and Lundqvist et al [8]. The cutoff parameter for the wave functions is 400 eV, and k-point meshes are generated through Monkhorst-Pack method with a spacing of $0.02 \times 2\pi$ Å$^{-1}$. The phonon calculations are performed by supercell approach implemented in the PHONOPY code [9] together with VASP as the atomic force calculator, adopting a $2 \times 2 \times 2$ supercell for 1T' and 2H structure. Electronic structure calculations are performed using full-potential linearized augmented plane-wave (FP-LAPW) method implemented in the WIEN2k [10] package. A 2000 k-point mesh for BZ sampling and 7 for the plane wave cut-off parameter $R_{MT}K_{max}$ are used in the calculation, where R is the minimum LAPW sphere radius and $K_{max}$ is the plane-wave vector cutoff. Spin-orbit coupling is taken into consideration in the second-variational calculation [11]. Transition barriers are calculated through the variable cell NEB (VC-NEB) method [12] implemented in the USPEX code [13]. The electron-phonon coupling (EPC) calculations are performed through the Quantum-ESPRESSO program [14], with a $6 \times 8 \times 4$ k-point mesh for EPC matrix and a $3 \times 4 \times 2$ q-point mesh for Dynamical matrix. The cutoffs are 60 Ry for the wave functions and 600 Ry for the charge density.

**High pressure powder X-ray diffraction (XRD) and Raman experiments.** A symmetric diamond anvil cell (DAC) with a pair of diamond culet size 400 µm was used for the high pressure Raman experiments. Re gasket, initially ~250 um thick, was pre-indented to ~40 um, and drilled at the very center to form a diameter of 200 µm for the sample chamber. A piece of WTe$_2$ sample purchased from 2D Semiconductors was cut into ~20 µm disk and was then placed near the center of the sample chamber, along with a couple of ruby spheres used as the pressure indicator. Ar pressure medium was loaded with the sample in the chamber using gas loading system in the Mineral Physics Lab of the University of Texas at Austin. Raman and ruby spectra were collected using Renishaw inVia Raman spectroscopy system equipped with a 532 nm green laser and 3000 line/mm grating. In order to prevent the sample from thermal damaging, laser power was restricted to <5 mW with integration time up to 5 minutes. Obtained spectral data were then analyzed according to background spectra, and peak position and FWHM extracted using Lorentzian fitting. Samples for XRD experiments were purchased from HQ Graphene, and ground using mortar to form randomly oriented powder. Externally heated DAC (EHDAC) with culet size 500 µm was used for high pressure XRD measurements. TZM gasket was pre-indented down to 37 µm thickness, and was drilled with 180 µm hole in the center. Powder $WTe_2$ were compressed to disks and stack-loaded in the sample chamber, in order to guarantee sufficient thickness and random orientation. Au particle and ruby were also loaded as pressure indicators. Ne gas was used for pressure transmitting medium. XRD experiments were conducted at GSECARS 13IDD beamline of the Advanced Photon Source, Argonne National Laboratory. An incident X-ray beamtime of approximately 0.3344 Å in wavelength and 2-3 µm in beamsize (FWHM) was used for the experiments, while X-ray diffraction patterns of the sample were collected by a MAR CCD detector. High temperature was applied either by external resistive heating or laser heating.

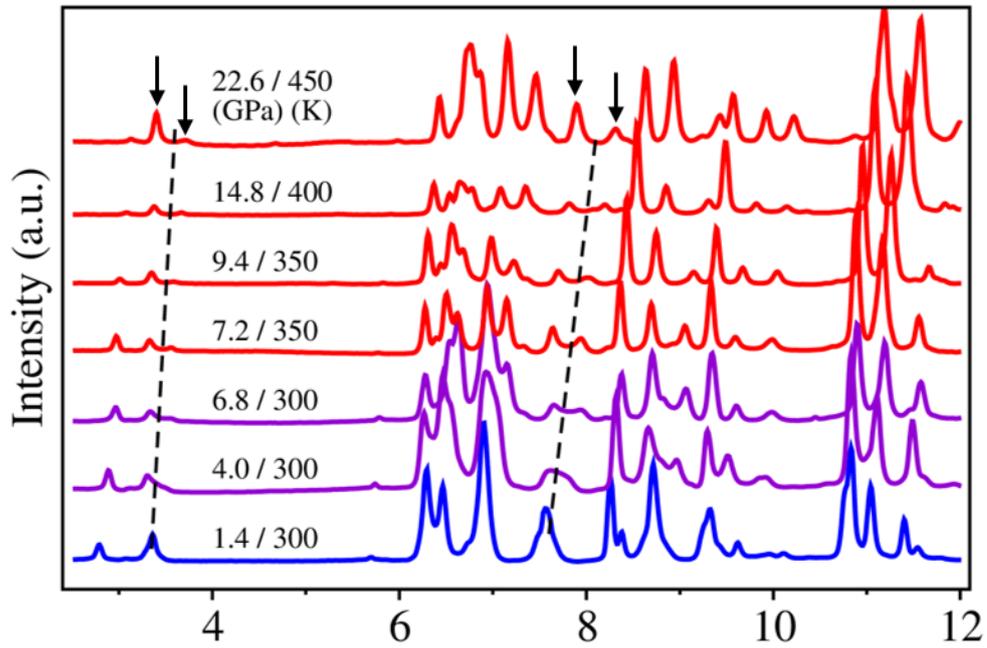

Figure S1. High pressure X-ray powder diffraction spectra measured under different pressures and temperatures. The (011) and (113) peak splitting in 1T' phase are observed, which can distinguish 1T' phase from Td phase. Dashed lines are guide to (011) peak and (113) peak, respectively

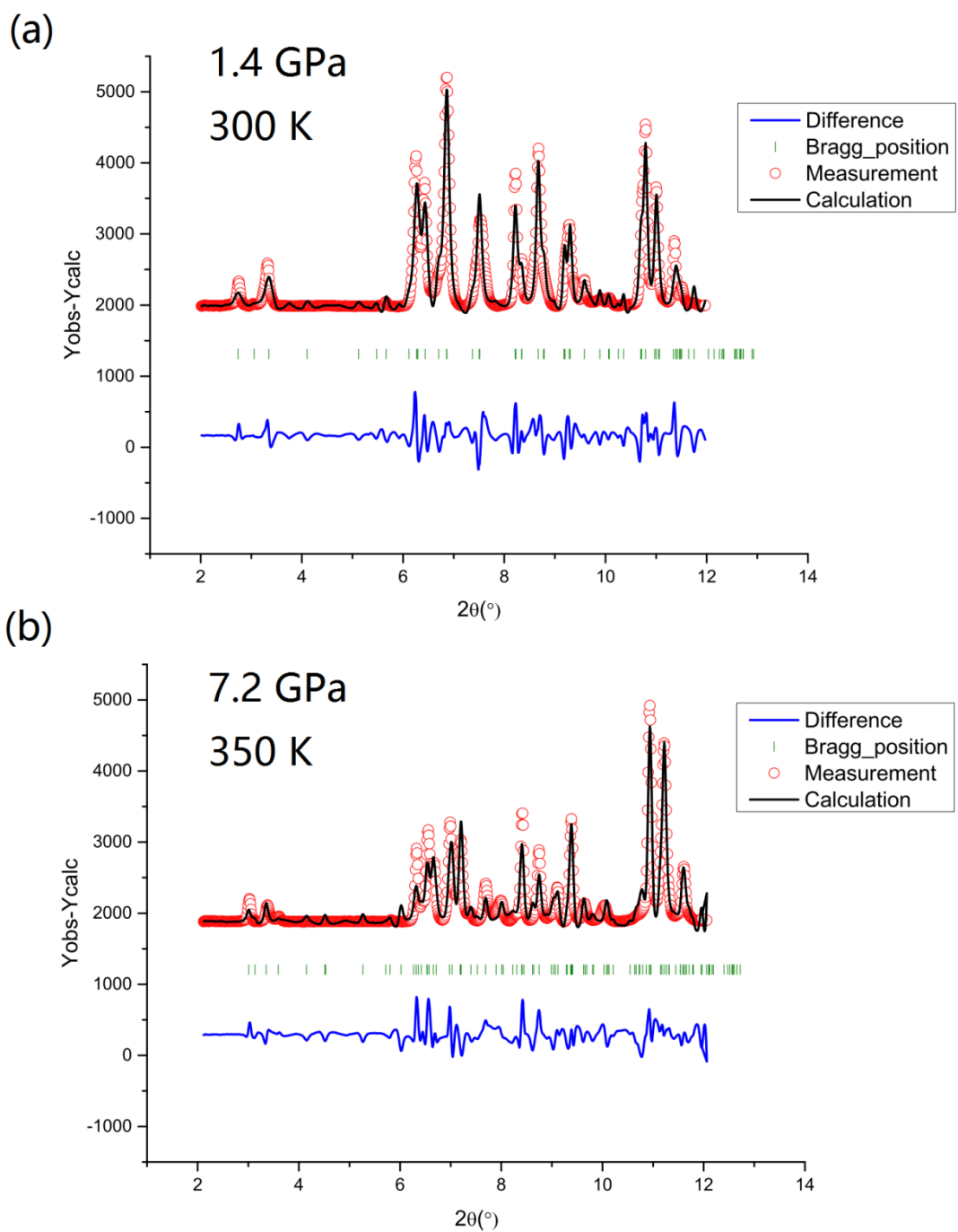

Figure S2. High pressure X-ray powder diffraction spectra for (a) Td phase 1.4 GPa / 300 K, and (b) 1T' phase 7.2 GPa / 350 K. The two experimental data matched well with calculated Bragg peaks.

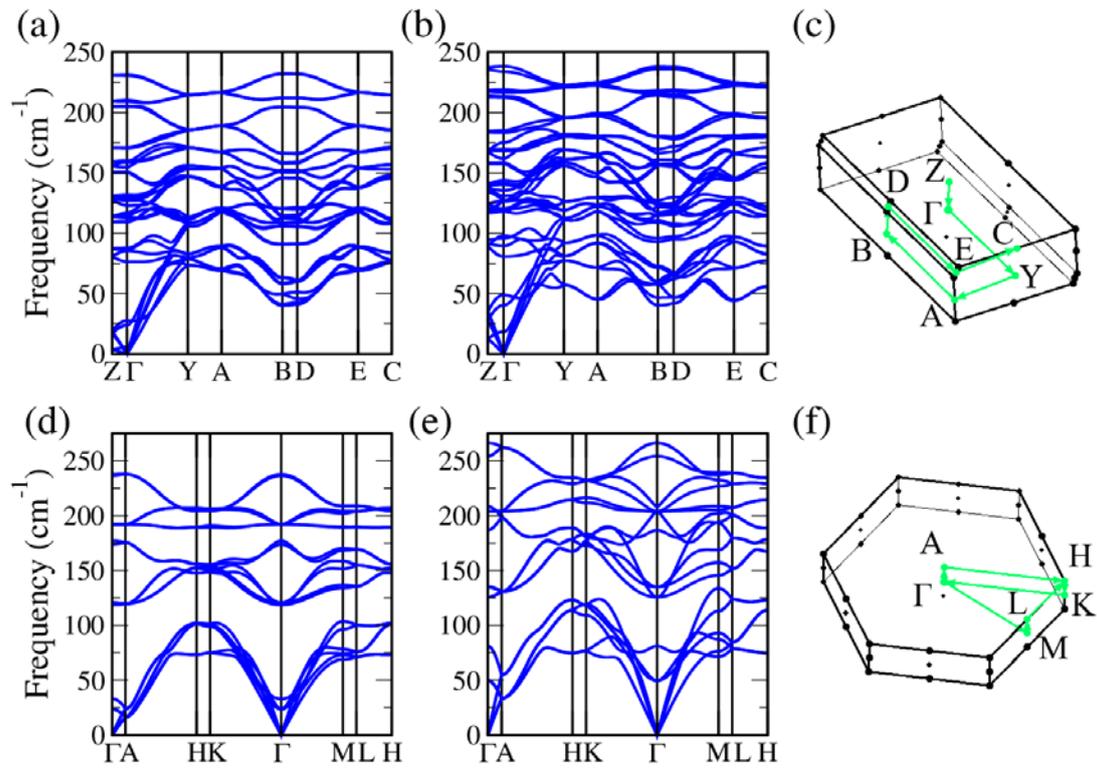

Figure S3. Phonon dispersion relations of 1T'-WTe$_2$ (a) 0GPa and (b) 5GPa, and 2H-WTe$_2$ at (d) 0GPa and (e) 15GPa, with their first Brillouin zones (c) and (f) respectively. These two structures are confirmed to be dynamically stable as the phonon dispersions exhibit no negative-frequency modes. moreover, it should be recoverable in decompression to ambient pressure if they could be synthesized experimentally. Two phonon modes near the A, C and E point goes soft in 1T'-WTe2 at 5GPa.

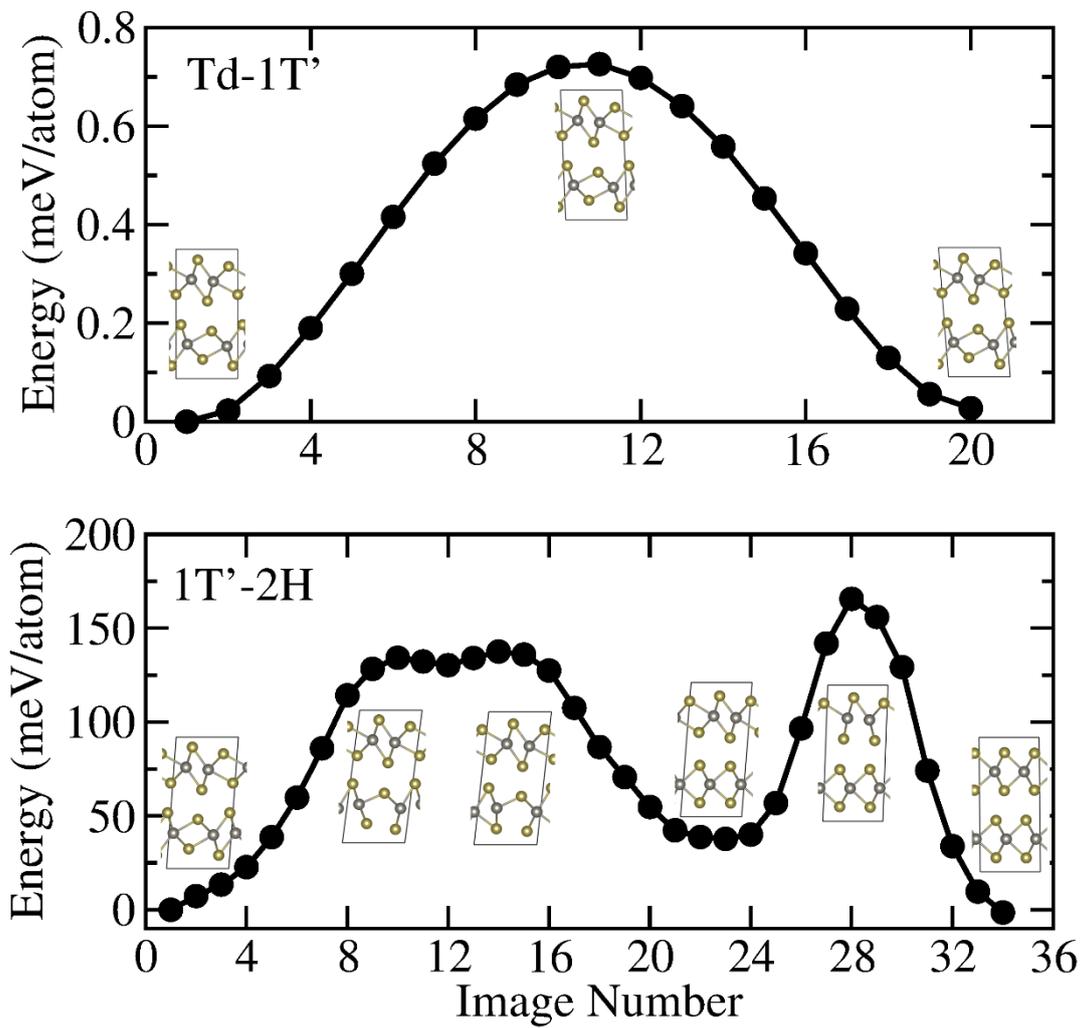

Figure S4. Enthalpy versus pathway for (a) Td-1T' transition at 5GPa and (b) 1T'-2H transition at 10GPa. The Td-1T' transition path is considered as shear strain along b axis, while the 1T'-2H transition path is considered as Te atomic layers gliding along b axis, with four bonds broken and rebuilt per cell.

Table S1. Calculated lattice parameters and volumes at 0GPa through GGA-PBE functional and optB88 functional with vdW-DF correction, compared with the experimental values from Ref. [15]. Normal GGA-PBE functional gives quite large error as compared to the experimental values, while the calculations based on optB88-vdW functional, gives error less than 1.0%.

|        | Td (Pmn21) | | | 1T' (P21/m) | | 2H (P63/mmc) | |
| --- | --- | --- | --- | --- | --- | --- | --- |
|        | PBE | optB88-vdW | Expt. | PBE | optB88-vdW | PBE | optB88-vdW |
| a      | 3.500 | 3.499 | 3.496 | 3.500 | 3.499 | 3.558 | 3.533 |
| b      | 6.328 | 6.285 | 6.282 | 6.327 | 6.286 | 3.558 | 3.533 |
| c      | 15.803 | 14.211 | 14.070 | 15.612 | 14.217 | 15.167 | 14.110 |
| Volume | 350.00 | 312.52 | 309.00 | 345.48 | 312.42 | 166.26 | 152.56 |